\documentstyle[aps,epsf,rotate,multicol]{revtex}

\begin{document}

\draft

\title{Environmental Changes, Co-extinction, and
Patterns in the Fossil Record}

\author{Lu\'{\i}s A. Nunes Amaral$^{\dagger}$ and Martin Meyer$^{*}$}

\address{$^{\dagger}$ Department of Physics, Massachusetts Institute 
	of Technology, Cambridge, MA 02139, USA \\
	 $^{*}$ Center for Polymer Studies and Dept. of Physics,
	Boston University, Boston, MA 02215, USA }


\maketitle

\begin{abstract}

We introduce a new model for large scale evolution and extinction in
which species are organized into food chains.  The system evolves by two
processes: origination/speciation and extinction.  In the model,
extinction of a given species can be due to an externally induced change
in the environment or due to the extinction of all preys of that species
(co-extinction).  The model is able to reproduce the empirical
observations, such as the statistical fractality of the fossil record or
the scale-free distribution of extinction events, without invoking
extinctions due to competition between species.

\end{abstract}

\begin{multicols}{2}

The identification of the mechanisms responsible for large-scale
evolution and extinction is a topic of heated debate
\cite{Gould77,Stanley79,Glen84,MSmith89,Elliot86,Raup93,Benton93,
Benton95,Sole97,Smith98}. The basic problem can be summarized by two
questions. The first one centers on the cause of mass extinction: Is
it external to the system --- e.g., due to extraterrestrial impacts
\cite{Alvarez80,Hoffmann91,Jablonski91,Raup94} --- or is it internal
and due to the nonlinear dynamics of the ecosystem
\cite{Kauffman91,Bak93,Sole96b,Newman96}?  The second question centers
on the mechanisms for selecting the species that become extinct: In
standard extrapolation of Darwinian evolution theory, it is assumed
competition leads to the extinction of less fit species
\cite{Kauffman91,Bak93,Sole96b,Newman96,Newman96b,Wilke97}, but some
authors argue that competition might be not a determinant factor for
macroevolution \cite{Gould77,Stanley79,Raup93}.  The fossil record has
yet to answer these questions unequivocally
\cite{Glen84,Signor82,House89,Raup91,Jablonski94}.  In qualitative
modelling, all of the mechanisms discussed above have been considered
as the explanation of the patterns in the fossil record.  On the other
hand, quantitative modelling approaches
\cite{Kauffman91,Bak93,Sole96b,Newman96} have consistently included
competition among species as a fundamental mechanism.  Here, we show
that a quantitative model that does not include competition among
species may reproduce the empirical observations, particularly the
statistical fractality of the fossil record \cite{Sole97,Burlando93},
and the scale-free distribution of extinction sizes
\cite{Elliot86,Raup93,Benton93}.

  The literature on large scale species extinction reports on two key
empirical results.  First, the probability density that a number $s$
of species becomes extinct during a given time interval decays as a
power law, $P(s) \sim s^{-\tau}$, with an exponent $\tau \approx 2$
\cite{Sole96b,Newman96,Raup91}.  Second, the power spectrum $S(f)$ of
the time series of extinction sizes also decays as a power law, $S(f)
\sim f^{-\beta}$, with $\beta \approx 1$ \cite{Sole97}, which implies
that the sequence of extinction is long-range correlated.  These
results impose severe constraints on the models attempting to describe
the extinction/evolution process.  A power law decay of the
probability of extinction sizes implies that there is no
characteristic size for extinction events, i.e. the dynamics are
scale-free and incidents of mass extinction are likely due to the same
mechanisms as smaller extinction events.  The hypothesis that the mass
extinctions are generated by the same dynamics as smaller extinction
events is consistent with the self-similarity of the fossil record
\cite{Sole97}.

  Quantitative models have been proposed to explain the patterns in
the fossil record.  Many are based on the assumption that extinction
events are a consequence of the competition between species, i.~e.~the
least fit species become extinct and are replaced by new species
\cite{Kauffman91,Bak93,Sole96b,Newman96,Newman96b,Wilke97}.  These
changes affect the fitness of other species leading to bursts of
extinction of all sizes.  Several of the models \cite{Bak93,Sole96b}
self-organize into a critical state in which many quantities are known
to scale as a power law \cite{Newman96}.  However, recently it has
been shown that mechanisms other than self-organized criticality, such
as coherent noise \cite{Newman96b,Wilke97} can lead to power law
scaling without requiring the system to be in a critical state.

\begin{figure}
\narrowtext
\centerline{
\epsfysize=0.8\columnwidth{\rotate[r]{\epsfbox{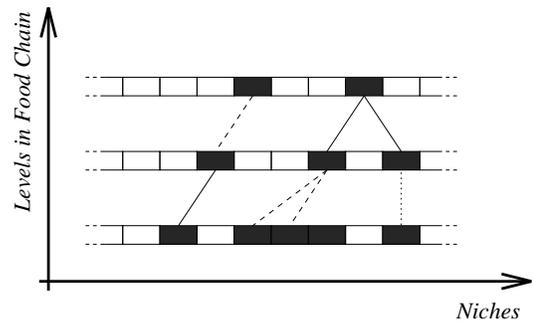}}}
}
\vspace*{1.0cm}
\caption{ Schematic definition of the model.  The evolution of the
	system takes place in a lattice in which each site represents
	a niche in the ``ecosystem''.  The system is organized into
	``trophic levels'', a species in level $\ell$ feeds from at
	most $k$ species in level $\ell - 1$, except for species at
	the first level which are autotrophic.  In most of the
	simulations there are 6 levels with 1000 niches per level.
	The state of the system is fully described by stating the
	niches which are occupied by a specie with the list of its
	preys.  We start the simulations with $N_o \approx 50$ species
	occupying niches in the first trophic level of the food chain.
	In the figure, the dark cells are occupied by a specie; the
	lines emerging from a cell link the species to its preys. The
	system evolves through two processes, origination and
	extinction.  {\bf Origination:} A niche in level $\ell$ is
	randomly selected, and if a species exist there, a speciation
	is attempted: A new niche is then randomly selected in one of
	the levels $\ell -1$, $\ell$, or $\ell + 1$, and if no species
	occupies that niche, a new species is created. {\bf
	Extinction:} A fraction $p$ of species in the first level are
	randomly selected for extinction. Then we remove for all
	species in the second level links to preys in the first level
	that have become extinct.  Whenever all links have been
	removed for a species in the second level, it becomes extinct
	as well. This procedure is repeated up the food chain until
	the top level is reached.  If, for the configuration in the
	figure, the leftmost species in the lowest level would become
	extinct, then the leftmost species in the other levels would
	also become extinct. }
\label{f.model}
\end{figure}

\begin{figure}
\narrowtext
\centerline{
\epsfysize=0.8\columnwidth{\rotate[r]{\epsfbox{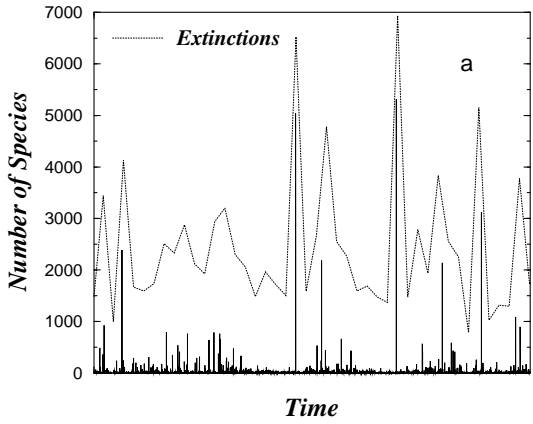}}}}
\vspace*{0.5cm}
\centerline{
\epsfysize=0.8\columnwidth{\rotate[r]{\epsfbox{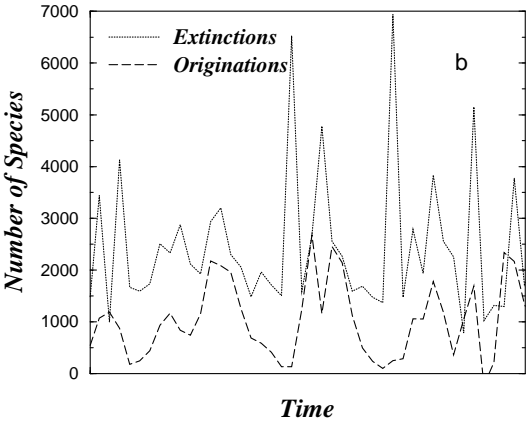}}}
}
\centerline{
\epsfysize=0.8\columnwidth{\rotate[r]{\epsfbox{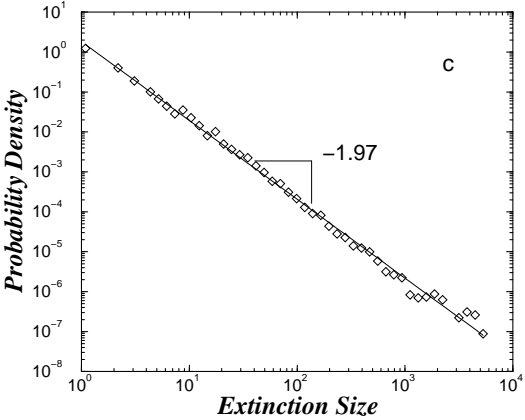}}}
}
\vspace*{1.0cm}
\caption{ Extinction events are scale-free. {\it a\/} Time sequence of
	extinction events for the model.  The lower line shows the
	individual events, while the upper curve shows the number of
	extinctions over a period of 512 time steps.  Note that events
	of all sizes (up to nearly the system size of 6000 species)
	are present.  The results shown are for a system with 6 levels
	and 1000 niches per level, a speciation rate of $\mu = 0.02$,
	and a extinction probability (due to environmental changes) of
	$p = 0.01$.  The results are only very weakly dependent on the
	values of the parameters.  {\it b\/} Time sequence of
	extinction and origination events. The origination curve is
	shifted downward by 1000 for clarity.  Note the strong
	correlation between the two curves, in agreement with
	empirical observations \protect\cite{Benton93,Holland97}.
	{\it c\/} Probability density function of events size.  The
	distribution is well described by a power law with an exponent
	$\tau = 1.97 \pm 0.05$, which is consistent with empirical
	measurements \protect\cite{Raup91}.  }
\label{f.analysis}
\end{figure}

\begin{figure}
\narrowtext
\centerline{
\epsfysize=0.8\columnwidth{\rotate[r]{\epsfbox{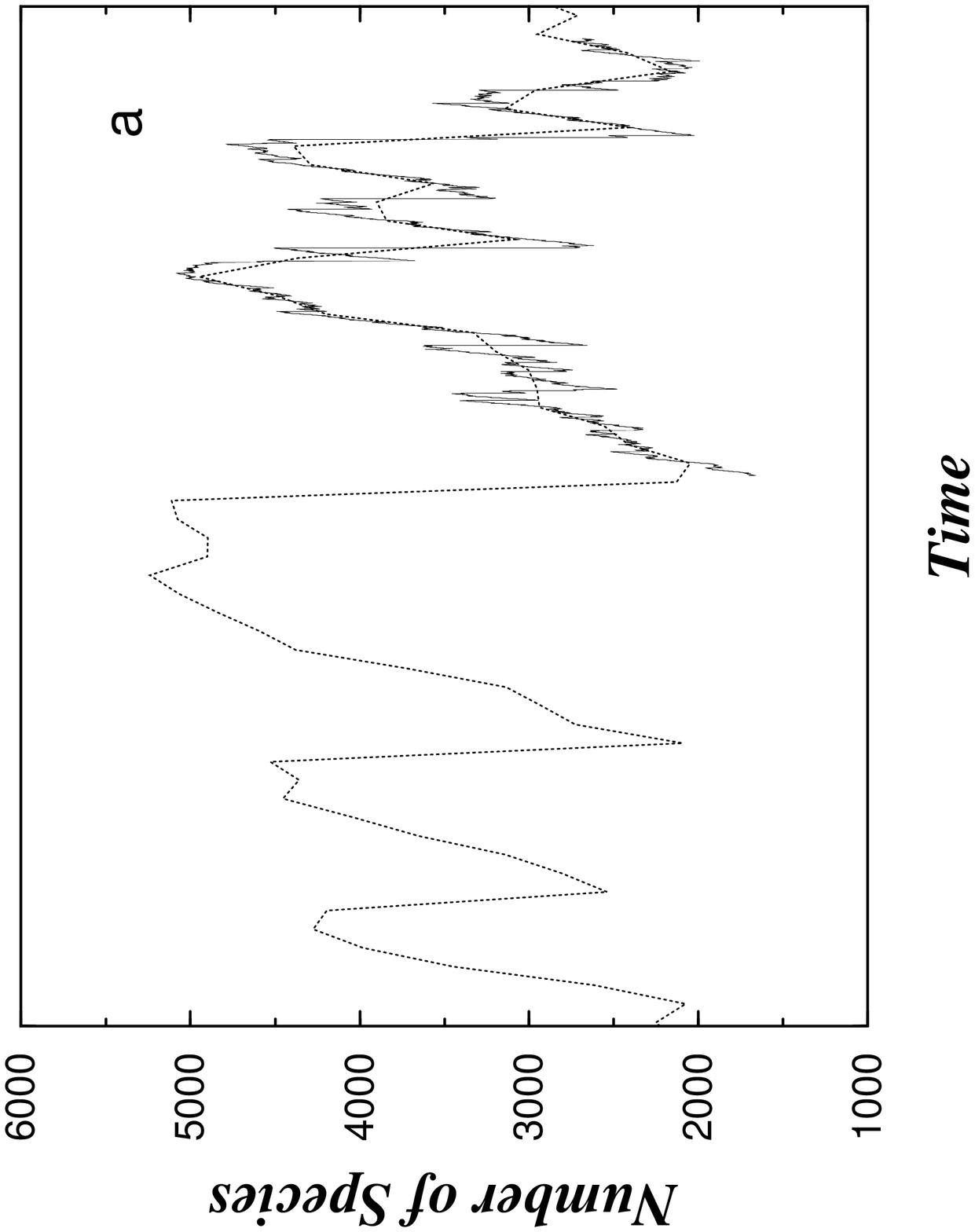}}}}
\vspace*{0.5cm}
\centerline{
\epsfysize=0.8\columnwidth{\rotate[r]{\epsfbox{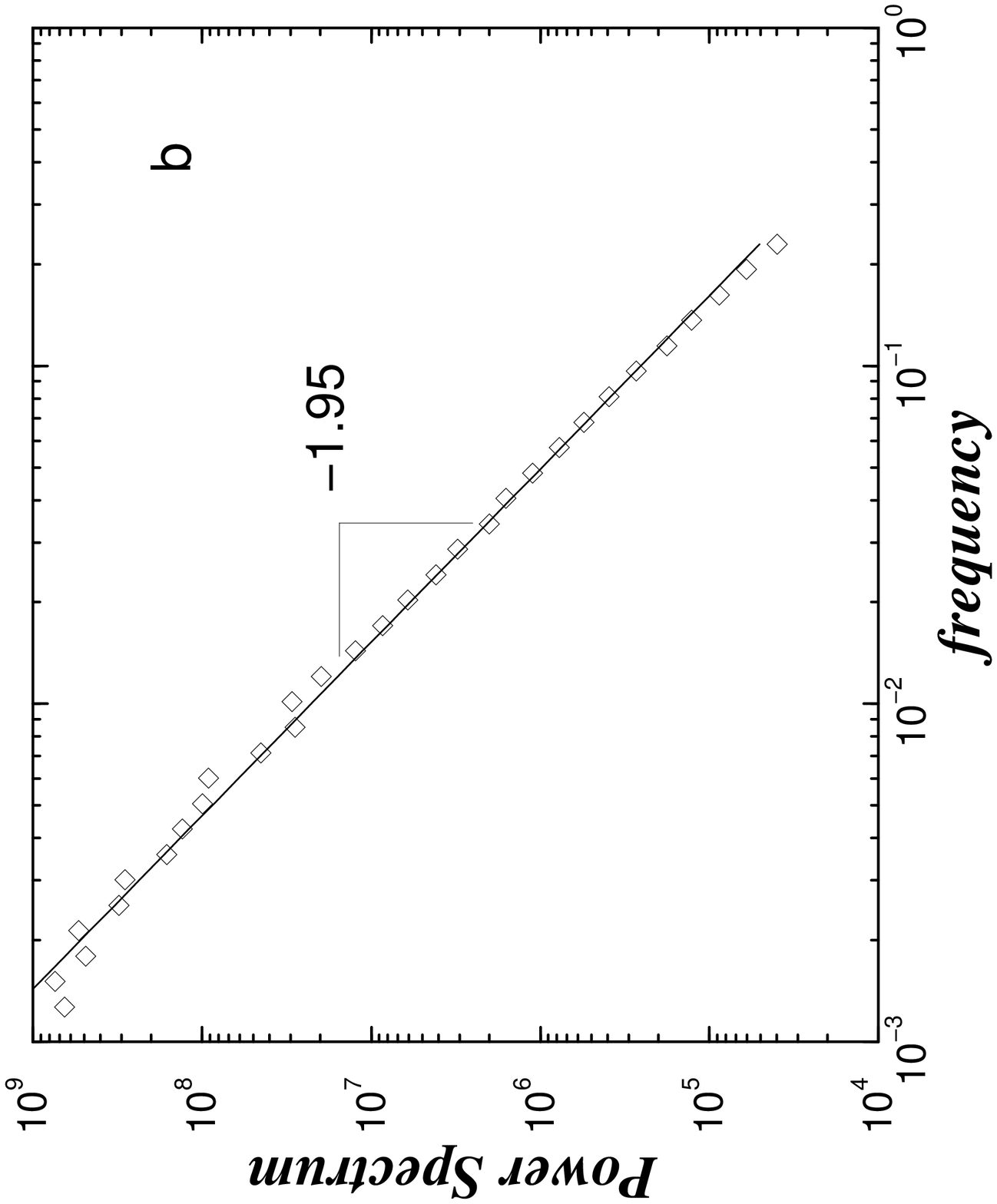}}}
}
\vspace*{1.0cm}
\caption{ Fractality of species diversity.  {\it a\/} Number of
	species in the model as a function of time.  The thicker
	dotted line shows the number of species at intervals of 128
	time steps.  The continuous line, shown for a shorter period
	of time, is sampled every time step.  Note the complex
	structure of the curve at very small time scales, which
	suggest that the fluctuations have a self-affine
	\protect\cite{Barabasi95} structure \protect\cite{Sole97}.
	{\it b\/} We investigate the power spectrum of the signal in
	{\it a\/} and find that it scales as a power law with an
	exponent $\beta = 1.95 \pm 0.05$, confirming the fractal
	nature of the fluctuations in the number of species for the
	model.  }
\label{f.diversity}
\end{figure}

  In this Letter, we test the hypothesis that competition between
species is {\it not} a fundamental ingredient for the explanation of
the fossil record. This hypothesis is in agreement with statements
that Darwinian competition while important at the level of individuals
within a population (microevolution) might not be relevant at the
level of stable species (macroevolution)
\cite{Gould77,Stanley79}. Thus, we propose a quantitative model for
large scale extinction and evolution that does not include competition
between species but assumes instead that the relevant mechanisms for
macroevolution are (random) changes in the environment
\cite{Hoffmann91,Jablonski91}, and co-extinctions \cite{Stork93} due
to the interactions between species along food chains
\cite{Smith98,Plotnick93,Sole96}.  The model is able to reproduce both
the power law distribution of extinction sizes and the fractality of
the fossil record [Figs.~\ref{f.model}--\ref{f.diversity}]. These
results suggest that {\it competition\/} between species might {\it
not\/} be a fundamental ingredient for the description of the fossil
record.

  The model is defined as follows.  Species can occupy niches in a
model ecosystem with $L$ trophic levels in the food chain, and $N$
niches in each level.  Species from the first level, $\ell = 0$ are
assumed to be autotrophic (i.e., they produce their food through,
e.g., photosynthesis), while species from levels $\ell > 0$ are
assumed to be heterotrophic.  That is, a species occupying a niche in
level $\ell > 0$ feeds from at most $k$ species occupying niches in
level $\ell-1$ [Fig.~\ref{f.model}].  We do not consider in the model
any kind of structure of the niches within a given trophic level, that
is, niches $i$ and $i+1$ in level $\ell$ do not need to be occupied by
similar species or to be geographically close.  Finally, we assume
that the preys of a new species are chosen at {\it random\/} from
existing species in the trophic level below.  The model starts with
$N_0$ species in level $\ell=0$ and evolves according to the following
rules:
\begin{itemize}
\item{} Origination: every existing species gives rise, at a rate
	$\mu$, to the creation of a new ``potential'' species that
	tries to occupy a randomly selected niche in the same trophic
	level or in one of the two neighboring levels.  This
	speciation event occurs if the selected niche is not yet
	occupied by an existing species.  Preys for the new species are
	selected at random from existing species in the trophic level
	below.
\item{} Extinction: at rate 1 (in some arbitrary time unit), a
  	fraction $p$ of species in the first level are randomly
  	selected for extinction.  Then, any species in the second
  	level for which all preys became extinct also becomes
  	extinct. This procedure is repeated up to level $L$.
\end{itemize}
These rules imply that the number of species in the system is not kept
constant.  In particular, if the origination rate is smaller than a
threshold value, then all species become extinct, i.e., the model has
absorbing states \cite{Grinstein}.  The rules for speciation imply
that the origination rate of new successful species is proportional to
the number $N_s$ of species in the system (leading to exponential
growth, in agreement with the results of \cite{Benton95}), and to the
number of empty niches $NL - N_s$ (which takes into consideration the
limited resources of the system \cite{Wilke97}).  Although the finite
size of the system introduces competition for the creation of new
species, the model does not involve any competition between existing
species.

  Figure~\ref{f.analysis} shows our results for the time sequence of
extinction and origination events.  The first interesting observation
is that both signals are intermittent with very large events appearing
at a high rate.  Furthermore, there is a strong correlation between
the extinction and origination curves, which is in qualitative
agreement with empirical observations \cite{Benton95,Holland97}.
Finally, we find that the size of the extinction events has a
distribution which decays with a power law tail with an exponent
$\tau = 1.97 \pm 0.05$, in agreement with empirical observations
\cite{Sole96b,Newman96,Raup91}.

  Next, we study the fluctuations in the number of species in the
system [Fig.~\ref{f.diversity}].  We find that these fluctuations are
self-affine \cite{Barabasi95}, as demonstrated by its power
spectrum which scales as a power law.  This result is in agreement
with the perceived fractality of the fossil record
\cite{Sole97,Burlando93}.

\begin{figure}
\narrowtext
\centerline{
\epsfysize=0.8\columnwidth{\rotate[r]{\epsfbox{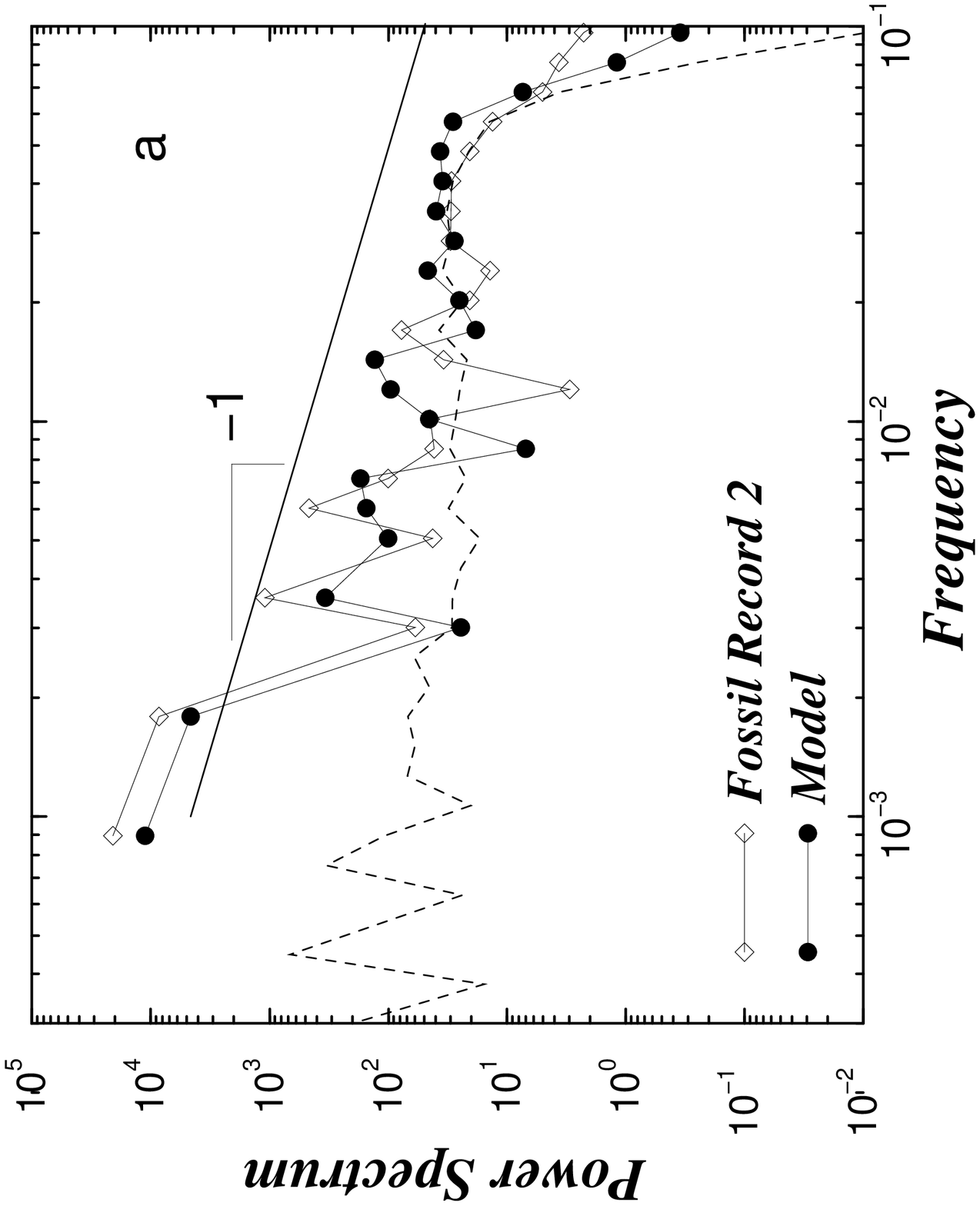}}}}
\vspace*{0.5cm}
\centerline{
\epsfysize=0.8\columnwidth{\rotate[r]{\epsfbox{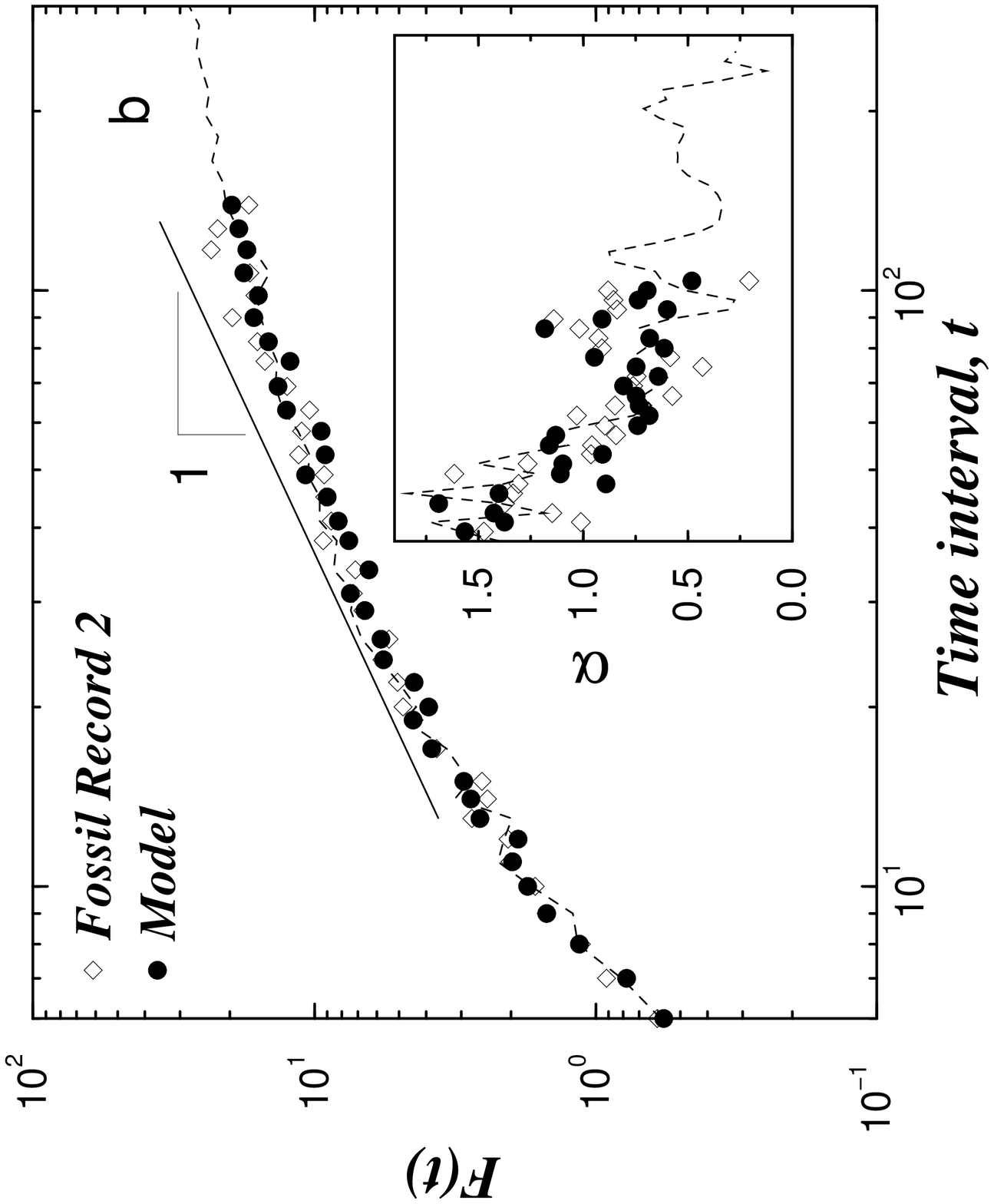}}}
}
\vspace*{1.0cm}
\caption{ Correlations in the fossil record and in the model. {\it
	a\/} We compare the results for the model with the empirical
	data found in \protect\cite{Benton93}.  For the model, we
	consider two sequences, one with 512 points (black circles)
	and another with 4096 points (dashed line).  The figure shows
	that the scaling behavior found for data and model is similar.
	We find that for about one order of magnitude the data for the
	shorter sequences appears to scale as a power law with an
	exponent $-1$.  However, it seems that such scaling does not
	hold for longer sequences, for which the power spectrum
	becomes flat, suggesting that the sequence crosses over to
	uncorrelated behavior (white noise). {\it b\/} We use
	detrended fluctuation analysis \protect\cite{Barabasi95} to
	test the results of the power spectrum.  We find $F(t)$, which
	measures fluctuations at different time scales, to scale as a
	power law with an exponent close to 1 for about one order of
	magnitude.  In the inset, we show the values of the exponent
	for a local fit to a power law.  Again all curves seem to
	behave in similar fashion.  However, the results suggest that
	no true scaling regime exist for time scales shorter than 300;
	then, the exponent becomes $1/2$ which would suggest an
	uncorrelated process. }
\label{f.nature.test}
\end{figure}

  In order to demonstrate the ability of our model to reproduce
quantitatively the empirical data on extinction and origination, we
compare in detail our results with the recent results of
Ref.~\cite{Sole97}. We therefore study the temporal correlations of
extinction events for the model and compare our results with the
analysis of the fossil record \cite{Benton93}.
Figure~\ref{f.nature.test} shows that the model results agree well
with the empirical data, when we consider model sequences of the same
lengths as available in the fossil record. This agreement is found for
the power spectrum as well as for the method of detrended fluctuation
analysis, which allows accurate estimates of correlation exponents
{\it independent} of local trends \cite{Barabasi95}. Note, however,
that once we consider longer records generated by the model, we find
that the results crossover to uncorrelated behavior. In fact, the
analysis of local slopes [see inset of Fig.~\ref{f.nature.test}b]
indicates a similar trend for the empirical data as well, suggesting
that extinction events might become uncorrelated at long time scales.

The model proposed here is able to reproduce key statistical
properties of the fossil record, both for the extinction and the
origination of species.  In contrast with many models in the
literature, these results are obtained {\it without\/} having to
assume that species have an intrinsic fitness, and that less fit
species become extinct due to competition between species.  In the
model, mass extinctions are due to the amplification effect of
predator-prey interactions that propagate along the food chain
\cite{Smith98}.  In this framework, the extinction of some key species
(due to environmental changes) can lead to catastrophic extinction
events.

We thank N.\ Dokholyan, P.Ch.\ Ivanov, H.\ Kallabis, M.\ Kardar, R.V.\
Sol\'e, and H.E.\ Stanley for stimulating discussions.  L.A.N.A.\ thanks
the JNICT and M.M.\ thanks the DGF for financial support.



\end{multicols}

\end{document}